\documentclass[a4paper]{jpconf}
\usepackage{graphicx}
\usepackage{amsmath}
\usepackage{amssymb}
\usepackage{dcolumn}
\usepackage{float}
\usepackage{tabularx}
\usepackage{xcolor}
\usepackage{enumerate}
\usepackage{bm}
\usepackage[square,sort&compress,numbers]{natbib}

\newcolumntype{d}[1]{D{.}{.}{#1}}
\newcommand{\iso}[2]{$^{#1}$#2}
\newcommand{\mev}[1]{$#1\,\mathrm{MeV}$}

\begin{document}
\title{Nuclear level density as a tool for probing the inelastic scattering of $\mathbf{{}^{6}He}$}

\author{Bora Canbula, Halil Babacan}

\address{Department of Physics, Faculty of Arts and Sciences,\\
Celal Bayar University, 45140 Muradiye, Manisa, Turkey}

\ead{bora.canbula@cbu.edu.tr}

\begin{abstract}
The cross sections are calculated for the both elastic and inelastic scattering of \iso{6}{He} from \iso{12}{C} and \iso{4}{He}. A phenomenological optical potential is used to describe the elastic scattering. \iso{4}{He} is taken as spherical and inelastic couplings to the first excited states of \iso{6}{He} and \iso{12}{C} are described with collective rotational model and coupled-channels method. Deformation lengths for \iso{6}{He} and \iso{12}{C} are determined from semi-classical nuclear level density model by using Laplace-like formula for the nuclear level density parameter. The comparison of the predicted and the measured cross sections are presented to test the applicability of nuclear level density model to the light exotic nuclei reactions. Good agreement is achieved between the predicted and measured cross sections.
\end{abstract}

\section{Introduction}
After the use of radioactive ion beams (RIBs) in the mid-1980s \cite{tanihata1985a}, heavy-ion induced reactions, which involve light exotic nuclei, have become a subject of great interest for both theoretical and experimental studies. Nevertheless, understanding the unusual structure of these nuclei, such as halo, skin or Borromean (three-body systems which have no bound two-body subsystems), still remains as a challenge. One of the most studied and also the lightest nucleus among them is \iso{6}{He}, which is 2n halo ($S_{2n}=0.975 \, \mathrm{MeV}$) \cite{tanihata1985b} and also Borromean ($\alpha+n+n$)\cite{bang1996}. Furthermore, \iso{6}{He} is of great astrophysical importance because of appearing to be a vital member of the reaction chain that can bridge the instability gaps at $A=5$ and $A=8$. 

Elastic scattering is the most fundamental and the simplest interaction between the projectile and target nuclei. Thus, at least for stable systems, it is easy to calculate the elastic cross section by using a simple optical potential. With this point of view, it would be a good starting point that employing the optical potential parameters deduced from the elastic scattering data of \iso{6}{Li}, for describing the elastic scattering of \iso{6}{He}. However, between the elastic scattering data of \iso{6}{Li} and \iso{6}{He}, there are certain differences such as long range absorption, absence of Coulomb rainbow, coupling to inelastic or breakup channels, which require unphysical modification of optical potential parameters to explain. Therefore, when it comes to explaining the elastic cross section of halo nuclei, some other mechanisms should be taken into consideration to deal with such phenomenon. Keeping in mind the low binding energy of \iso{6}{He}, there is no doubt that the CDCC (continuum-discretized coupled-channels) model is the most appropriate method, but it is still a good choice to combine the phenomenological optical potential with some virtual coupling or dynamic dipole polarization potential, which are surface (derivative) potentials of the usual volume potential, to describe the elastic scattering data of \iso{6}{He} reasonably well. 
\begin{figure}[t!]
\begin{center}
\begin{tabular}{cc}
\includegraphics{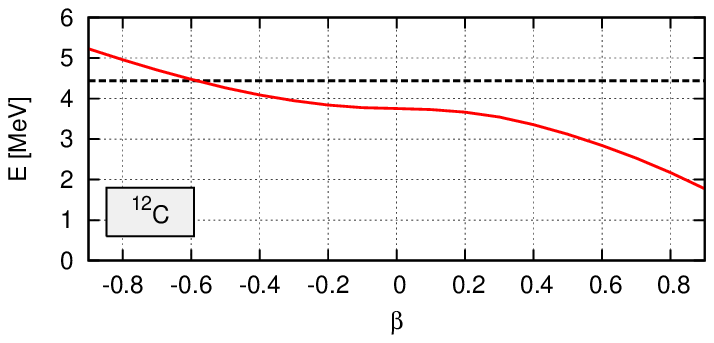}&\includegraphics{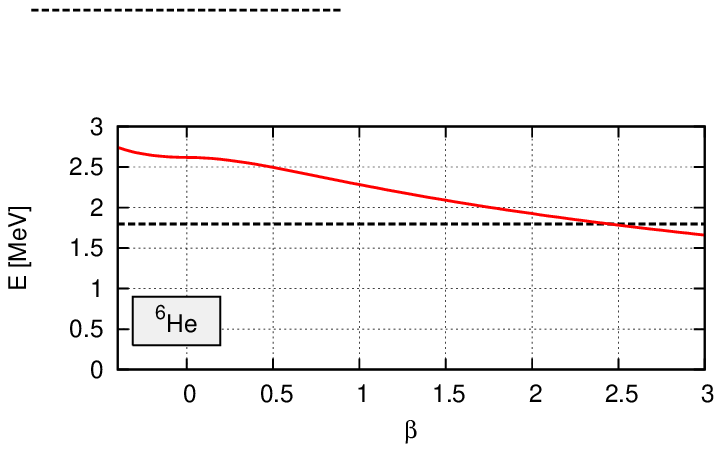}\\
\end{tabular}
\end{center}
\caption{\label{fig:figure1} (Color online) The energy of the first excited state as a function of deformation parameter. The results are illustrated for \iso{12}{C} and \iso{6}{He} in the left and right panels, respectively.}
\end{figure}

The main idea behind these theoretical efforts is to simulate the absorption from elastic channel due to the surface deformation of \iso{6}{He} compared to a stable $A=6$ system. Therefore, one can describe the absorption from elastic channel as inelastic couplings to the low-lying collective states by using coupled-channels method. To use this method the deformation lengths of the projectile and target should be determined from a structure model. Nuclear level density (NLD) could have been one of the candidates but it always has been a problem to describe the low-lying collective states for NLD models \cite{koning2008,canbula2011}. To overcome this problem, recently, we have proposed a Laplace-like formula for the energy dependence of the nuclear level density parameter, which is the main parameter of NLD, including collective effects \cite{canbula2014a}. Consequently, a significant improvement has been achieved in agreement between the predicted and the observed excited energy levels. In this model, the density of the excited levels are calculated in terms of deformation parameter of nucleus and also a semi-classical approach is employed with a single-particle potential. 
\begin{table}[b!]
\caption{\label{tab:table1}Adjusted optical potential parameters. $r_c=1.2\,\mathrm{fm}$ is used in all calculations.}
\begin{center}
\begin{scriptsize}
\begin{minipage}{.75\textwidth}
\begin{tabular}{llllllll}\br
Proj.+Targ. &$E(\mathrm{MeV})$&$V_0(\mathrm{MeV})$&$r_v\,(\mathrm{fm})$&$a_v\,(\mathrm{fm})$&$W_0\,(\mathrm{MeV})$&$r_w\,(\mathrm{fm})$&$a_w\,(\mathrm{fm})$\\
\mr
\end{tabular}\\
\begin{tabular}{ld{5.3}d{3.3}d{3.3}d{3.3}d{4.3}d{4.3}d{3.3}}
\iso{6}{He} + \iso{12}{C} & 18.0^{*} & 44.789 & 0.738 & 0.867 & 11.284 & 1.227 & 0.696 \\
\iso{6}{He} + \iso{12}{C} & 30.0^{*} & 48.775 & 0.931 & 0.745 &  6.991 & 1.310 & 0.890 \\
\iso{6}{He} + \iso{4}{He} & 3.8^{**} & 64.800 & 0.910 & 0.660 &  0.500 & 0.900 & 0.650 \\
\iso{6}{He} + \iso{4}{He} & 4.2^{**} & 64.000 & 0.910 & 0.660 &  2.200 & 0.900 & 0.650 \\
\iso{6}{He} + \iso{4}{He} & 4.7^{**} & 62.800 & 0.910 & 0.660 &  2.400 & 0.900 & 0.650 \\
\iso{6}{He} + \iso{4}{He} & 5.1^{**} & 62.000 & 0.910 & 0.660 &  2.600 & 0.900 & 0.650 \\
\iso{6}{He} + \iso{4}{He} & 5.4^{**} & 61.200 & 0.910 & 0.660 &  3.000 & 0.900 & 0.650 \\
\iso{6}{He} + \iso{4}{He} & 5.8^{**} & 60.800 & 0.910 & 0.660 &  3.200 & 0.900 & 0.650 \\
\br
\end{tabular}\\
\begin{footnotesize}
$^{*}$ Energy in lab. system\\
$^{**}$ Energy in c.m. system
\end{footnotesize}
\end{minipage}
\end{scriptsize}
\end{center}
\end{table}
\begin{figure}[t!]
\begin{center}
\includegraphics{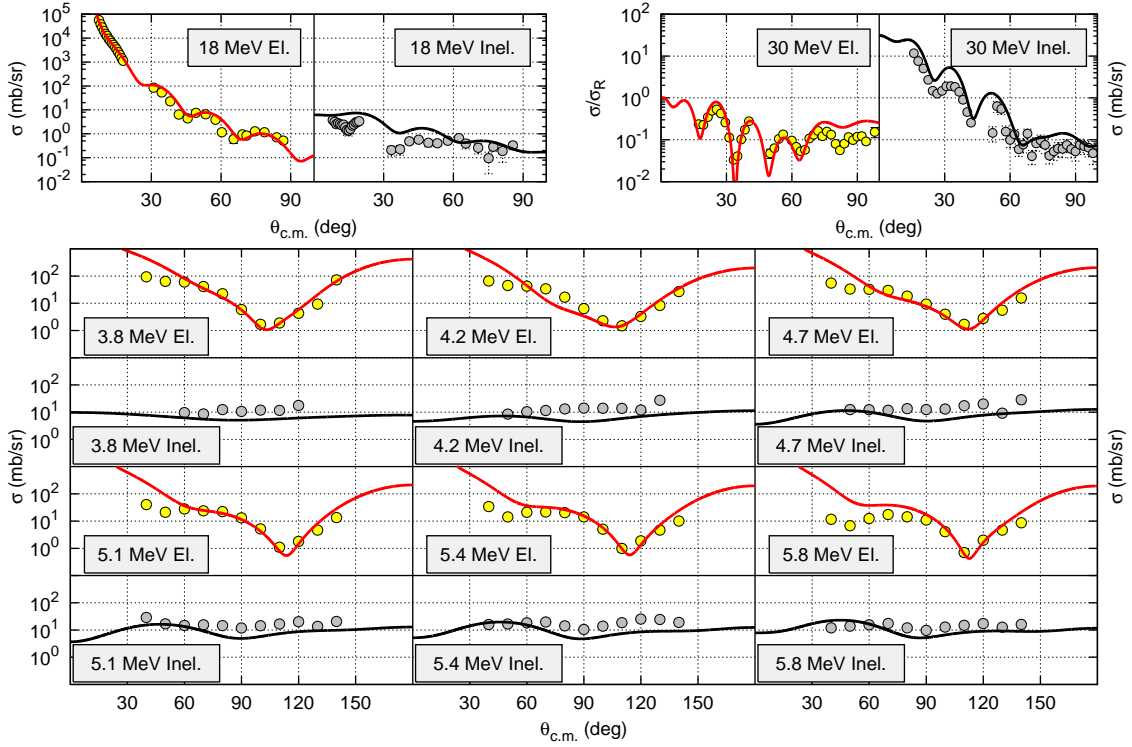}
\end{center}
\caption{\label{fig:figure2} (Color online) The results with comparison to experimental data \cite{milin2004,smalley2014,suzuki2013} for elastic and inelastic scattering of \iso{6}{He} on \iso{12}{C} (in the upper part) target at \mev{18} and \mev{30} laboratory energies and \iso{4}{He} (in the lower part) target at \mev{3.8}, \mev{4.2}, \mev{4.7}, \mev{5.1}, \mev{5.4} and \mev{5.8} center-of-mass energies.}
\end{figure}

\section{Nuclear level density}
The easiest way to determine the nuclear level density of a nucleus is to use a phenomenological model, and one of the most common phenomenological nuclear level density model is Fermi gas model. According to this model, the total nuclear level density is 
\begin{equation}
\label{eq:totalleveldensity}
\rho^{\mathrm{tot}}(U) = \frac{1}{12 \sqrt{2} \sigma} 
\frac{\mathrm{exp}[2 \sqrt{a U}]}{a^{1/4} U^{5/4}},
\end{equation}
where $U$ is effective excitation energy and $\sigma$ is spin cut-off parameter. The main variable of this equation is $a$, which is called as the nuclear level density parameter. In this study, we use the Laplace-like formula \cite{canbula2014a}, which is convenient to describe the low-lying collective levels, for the energy dependence of the nuclear level density parameter. This formula, 
\begin{equation}
\label{eq:ldp_canbula}
a(U) = \tilde{a} \left ( 1 + A_{c} \frac{S_{n}}{U} \frac{\exp (- | U - E_{0} | / {\sigma\prime}_{c}^{3})}{{\sigma\prime}_{c}^{3}} \right ),
\end{equation}
gives the energy dependence of the nuclear level density parameter with a function that depends on the neutron separation energy $S_{n}$, the energy of the first phonon state $E_{0}$, and a scale parameter, which is given as ${\sigma\prime}_{c}^{3}=\sigma_{c}^{3}/\tilde{a}$ in terms of spin cut-off parameter and the asymptotic level density parameter. The so-called collective amplitude $A_{c}$ is given as the temperature dependent difference between the experimental mass and the theoretical mass calculated with a shape dependent liquid-drop mass formula, 
\begin{equation}
\label{eq:collectiveamplitude}
A_{c} = \left [ M_{\mathrm{exp}}(N,Z)-M_{\mathrm{LDM}}(N,Z,\beta) \right ] \frac{\tau_{c}}{\sinh \tau_{c}},
\end{equation}
where $T_{c}=\sqrt{S_{n}/\tilde{a}}$ is defined as some critical temperature and $\tau_{c}=2\pi^{2}T_{c}/\hbar\omega$. Further details can be found in Ref. \citep{canbula2014a}. One can use the level density function to obtain the cumulative number of levels up to certain excitation energy,
\begin{equation}
\label{eq:cumulativelevels}
N_{\mathrm{cum}}(U) = \int_{0}^{U} \rho^{\mathrm{tot}}(U) dU.
\end{equation}

\section{Results and discussion}
In this study, we evaluated the integral \eqref{eq:cumulativelevels} with certain deformation parameter values and found the upper limit energy which gives $N_{\mathrm{cum}}=1$ to find the deformation parameter value of the nucleus for the first excited state. With this procedure we were able to obtain the energy of the first excited state as a function of deformation parameter. The obtained results for \iso{12}{C} and \iso{6}{He} are given in Figure \ref{fig:figure1}. In this figure, the red solid line represents the results of our calculation and the black dashed line corresponds the experimentally known value for the energy of the first excited state. As can be seen from the figure, our model predicts the deformation of \iso{12}{C} as $-0.6$ and \iso{6}{He} as $2.45$. 

To do calculations with the coupled-channels method for the inelastic couplings by using the obtained deformation parameters, the optical potential parameters should be adjusted to elastic scattering data first. Adjusted parameters are given in Table \ref{tab:table1}. We used the Woods-Saxon potential combined with a Coulomb term as optical potential. For \iso{6}{He}+\iso{12}{C} system, we found different geometrical parameters for \mev{18} and \mev{30}, but for \iso{6}{He}+\iso{4}{He} system we could describe the data at six different energies with the same geometrical parameter values and slowly varying potential depths with the energy. The results of our calculations for these two systems at eight different energies are shown in Figure \ref{fig:figure2}. Good agreement between the predicted and measured cross sections is achieved as seen from figure.

\section{Conclusions}
Theoretical calculations are performed for both elastic and inelastic scattering of \iso{6}{He} from \iso{12}{C} and \iso{4}{He}. These preliminary results, which are given in this paper, show that the nuclear level density can be used as a tool for calculations of inelastic scattering. Obtained deformation parameter of \iso{12}{C} is in agree with the existing literature \cite{nunes1996,vermeer1983}. Although the sign of the predicted deformation parameter for \iso{6}{He} is consistent with some other studies \cite{lee2007}, because of its large magnitude it should be reconsidered and a more complicated shape definition should be used to calculate the collective amplitudes. Further details of this study will be published elsewhere.

\ack
This work was supported by the Turkish Science and Research Council (T\"{U}B\.{I}TAK) under Grant No. 112T566. Bora Canbula acknowledges the support through T\"{U}B\.{I}TAK PhD Program fellowship B\.{I}DEB-2211 Grant.

\bibliography{Canbula}
\end{document}